\documentclass{PoS}

\title{ThunderKAT: The MeerKAT Large Survey Project for Image-Plane Radio Transients}

\ShortTitle{ThunderKAT}

\author{\speaker{Rob Fender$^{1,2*}$, Patrick Woudt$^2$}\\
  E-mail: \email{rob.fender@physics.ox.ac.uk, Patrick.Woudt@uct.ac.za}}

\author{Richard Armstrong$^{1,2}$,
    Paul Groot$^3$, Vanessa McBride$^{2,4}$, James Miller-Jones$^5$, Kunal Mooley$^{1}$,
    Ben Stappers$^6$, Ralph Wijers$^7$}

\author{Michael Bietenholz$^{8,9}$, Sarah Blyth$^2$, Markus Bottcher$^{10}$, David Buckley$^{4,11}$, Phil Charles$^{1,12}$,
    Laura Chomiuk$^{13}$, Deanne Coppejans$^{14}$, St\'ephane Corbel$^{15,16}$, Mickael Coriat$^{17}$, Frederic Daigne$^{18}$, Erwin de Blok$^2$,
    Heino Falcke$^3$, Julien Girard$^{15}$, Ian Heywood$^{19}$, Assaf Horesh$^{20}$, Jasper Horrell$^{21}$,
    Peter Jonker$^{3,22}$, Tana Joseph$^4$, Atish Kamble$^{23}$, Christian Knigge$^{12}$, Elmar K\"ording$^3$, Marissa Kotze$^4$,
    Chryssa Kouveliotou$^{24}$, Christine Lynch$^{25}$, Tom Maccarone$^{26}$, Pieter Meintjes$^{27}$, Simone Migliari$^{28}$,
    Tara Murphy$^{25}$, Takahiro Nagayama$^{29}$, Gijs Nelemans$^3$, George Nicholson$^8$, Tim O'Brien$^6$,
    Alida Oodendaal$^{27}$, Nadeem Oozeer$^{21}$, Julian Osborne$^{30}$, Miguel Perez-Torres$^{31}$, Simon Ratcliffe$^{21}$,
    Valerio Ribeiro$^{32}$, Evert Rol$^6$, Anthony Rushton$^1$, Anna Scaife$^6$, Matthew Schurch$^2$, Greg Sivakoff$^{33}$,
    Tim Staley$^1$, Danny Steeghs$^{34}$, Ian Stewart$^{35}$, John Swinbank$^{36}$, Kurt van der Heyden$^2$,
    Alexander van der Horst$^{24}$, Brian van Soelen$^{27}$, Susanna Vergani$^{37}$, Brian Warner$^2$, Klaas Wiersema$^{30}$\\
    \ \\
    $^1$ Astrophysics, Department of Physics, University of Oxford, UK \\
    $^2$ Department of Astronomy, University of Cape Town, South Africa\\
    $^3$ Department of Astrophysics, Radboud University, Nijmegen, the Netherlands\\
    $^4$ South African Astronomical Observatory, Cape Town, South Africa\\
    $^5$ ICRAR, Curtin University, Perth, Australia\\
    $^6$ University of Manchester, UK\\
    $^7$ Anton Pannekoek Institute for Astronomy, University of Amsterdam, the Netherlands\\
    $^8$ Hartebeeshoek Radio Observatory, South Africa \\
    $^9$ Department of Physics and Astronomy, York University, Toronto, Canada \\
    $^{10}$ North-West University, Potchefstroom campus, South Africa\\
    $^{11}$ Southern African Large Telescope, Cape Town, South Africa \\
    $^{12}$ School of Physics and Astronomy, Southampton University, UK \\
    $^{13}$ Department of Physics and Astronomy, Michigan State University, USA \\
    $^{14}$ Northwestern University, USA \\
    $^{15}$ Laboratoire AIM (CEA/IRFU - CNRS/INSU - Univ. Paris Diderot), Paris, France \\
    $^{16}$ Station de Radioastronomie de Nancay, Observatoir de Paris, France \\
    $^{17}$ University of Toulouse, UPS-OMP, France }

  \author{ \\
    $^{18}$ Institute d'Astrophysique de Paris, France \\
    $^{19}$ CSIRO Astronomy and Space Science, Epping, Australia \\
    $^{20}$ Benoziyo Center fo Astrophysics, Weizmann Institute of Science, Israel \\
    $^{21}$ SKA South Africa, Pinelands, South Africa \\
    $^{22}$ SRON, Netherlands Institute for Space Research, Utrecht, the Netherlands \\
    $^{23}$ Harvard-Smithsonian Center for Astrophysics, Cambridge, USA\\
    $^{24}$ Department of Physics, The George Washington University, Washington USA \\
    $^{25}$ School of Physics, University of Sydney, Australia \\
    $^{26}$ Department of Physics, Texas Tech University, USA \\
    $^{27}$ Department of Physics, University of the Free State, Bloemfontein, South Africa \\
    $^{28}$ European Space Agency, Madrid, Spain \\
    $^{29}$ Department of Physics and Astronomy, University of Kagoshima, Japan \\
    $^{30}$ Department of Physics and Astronomy, University of Leicester, UK \\
    $^{31}$ Insituto de Astrofisica de Andalucia, Spain \\
    $^{32}$ Botswana International University of Science and Technology, Botswana \\
    $^{33}$ Department of Physics, University of Alberta, Canada \\
    $^{34}$ Department of Physics, University of Warwick, UK \\
    $^{35}$ Sterrewacht Leiden, University of Leiden, the Netherlands \\
    $^{36}$ Department of Astrophysical Sciences, Princeton University, USA \\
    $^{37}$ GEPI, Observatoire de Paris, France }

\abstract{ThunderKAT is the image-plane transients programme for MeerKAT.
  The goal as outlined in 2010, and still today, is to find, identify and understand high-energy
  astrophysical processes via their radio emission (often in concert with observations at other
  wavelengths). Through a comprehensive and complementary programme of surveying and monitoring
  Galactic synchrotron transients (across a range of compact accretors and a range of other
  explosive phenomena) and exploring distinct populations of extragalactic synchrotron transients
  (microquasars, supernovae and possibly yet unknown transient phenomena) - both from
  direct surveys and commensal observations - we will revolutionise our understanding of the
  dynamic and explosive transient radio sky. As well as performing targeted programmes of our own,
  we have made agreements with the other MeerKAT large survey projects (LSPs) that we will also search their data
  for transients. This commensal use of the other surveys, which remains one of our key programme goals
  in 2016, means that the combined MeerKAT LSPs will produce by far the largest GHz-frequency radio transient
  programme to date.}

\FullConference{MeerKAT Science: On the Pathway to the SKA,\\
	25-27 May, 2016,\\
		Stellenbosch, South Africa}

\begin{document}

\section{Introduction}

Transient radio emission is associated with essentially all explosive phenomena and high-energy astrophysics in the universe. It acts as a locator for such events, and a calorimeter of their kinetic feedback to the local environment. Because of this, radio transients are invaluable probes for subjects as diverse as stellar
evolution, relativistic astrophysics and cosmology. ThunderKAT is the programme to observe such phenomena in the image plane with MeerKAT. It is also the umbrella project behind the MeerLICHT optical telescope \cite{2016SPIE.9906E..64B}, which will provide simultaneous optical images for all night-time MeerKAT observations.

\section{Accretion physics}

Accretion phenomena on all scales and across all environments are associated with the generation of outflows, which reveal themselves via synchrotron radiation in the radio band. Observations of these outflows allow us to test how much kinetic energy is being fed back from the accretion process, and in some cases to directly resolve the fastest-moving bulk phenomena in the universe.

\subsection{Relativistic accretion}

\subsubsection{Black holes and neutron stars in X-ray binaries}

Stellar-mass black holes (BHs) in X-ray binary systems provide our best
sample with which to understand how varying accretion states affect the
power and type of kinetic feedback from BH via winds and jets. The
neutron star (NS) X-ray binaries provide the best possible 'control sample'
to study which aspects of this behaviour are unique to BH and may therefore
be dependent on the existence of an event horizon or ergosphere.

\begin{figure}
\includegraphics[width=1.0\textwidth]{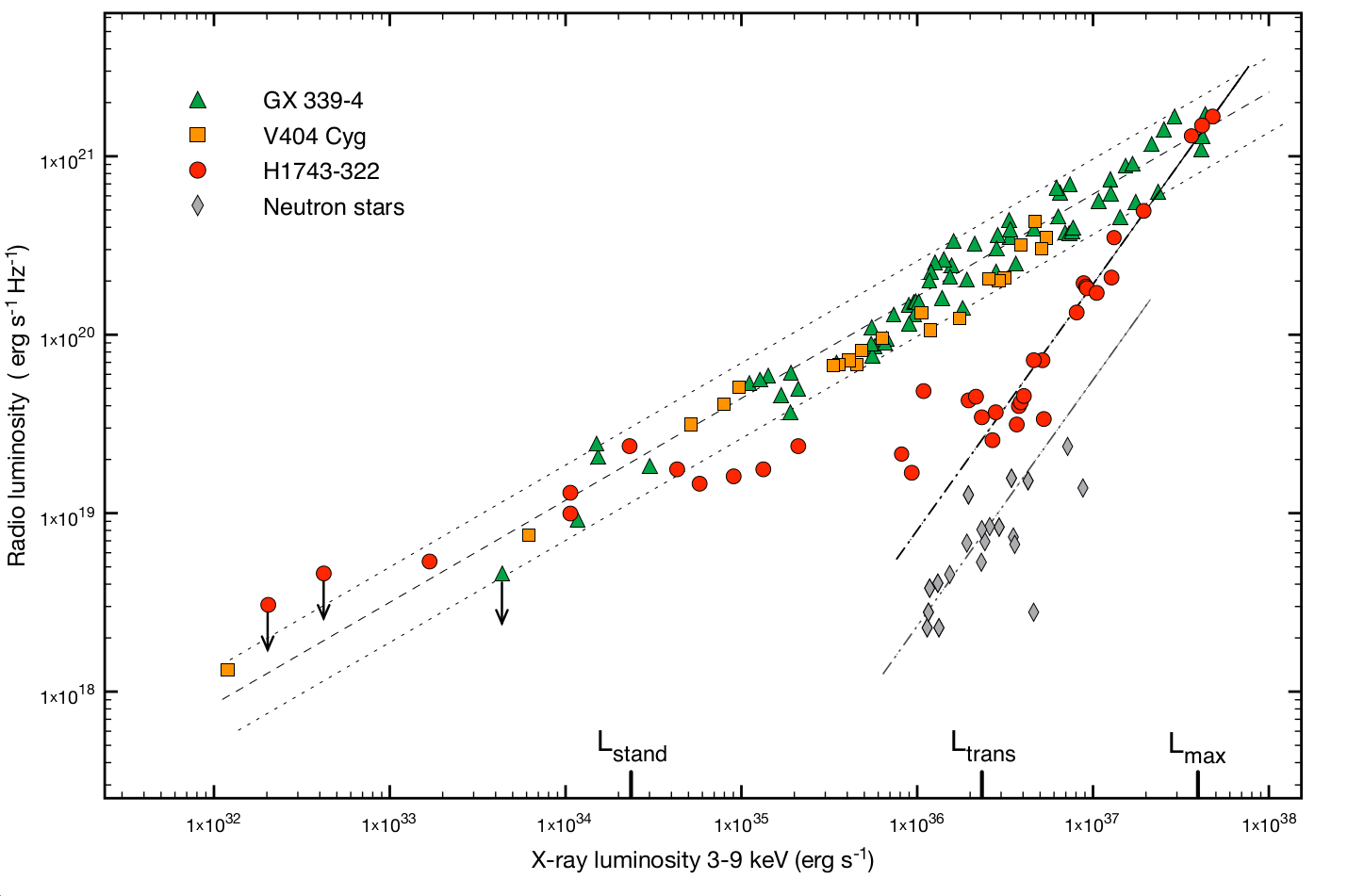}
\caption{The radio-X-ray plane for black hole and neutron star X-ray binaries \cite{2011MNRAS.414..677C}.
  Quasi-simultaneous observations allow us to map out the long-term connection between accretion and ejection.
  The empirical patterns observed can be scaled to supermassive black holes in AGN, and yet throw up many mysteries.
  A comprehensive monitoring programme with MeerKAT will increase the number of measurements in this plane by a
  factor of up to 5, notably probing for the first time the mapping for the neutron star binaries, which
  are radio-quiet relative to the black holes.}
\label{fig1}
\end{figure}

The current fundamental tool for understanding the global connection
between accretion and ejection in X-ray binaries is the radio-X-ray plane
(Fig 1). Currently the plane is populated by around 200
quasi-simultaneous radio and X-ray observations of X-ray binaries and
has led to some key results and even more unanswered questions:

\begin{itemize}
\item{a strong correlation exists between accretion rate and jet power, which
  can be extended with the addition of a mass term to include BH of all masses,
  from X-ray binaries to supermassive black holes in AGN.}
\item{the jet is suppressed in some high-luminosity states (analogues of radio-quiet AGN)}
\item{some neutron stars (the transitional millisecond pulsars) behave like black
  holes in this plane, whereas others show the steeper correlation expected for
  radiatively efficient accretion in the presence of a hard surface}
\item{there is a mysterious 'radio quiet' branch where some BH produce less powerful
  jets under the same accretion conditions (as measured in X-rays) as others.}
\end{itemize}

Note that for sources in bright accretion modes, X-ray monitors (e.g. ASTROSAT, HXMT)
will provide good enough measurements to place the MeerKAT data in the plane.
For fainter sources we will request e.g. Swift/Chandra/XMM observations
quasi-simultaneously with MeerKAT. Because the neutron star X-ray binaries
are significantly fainter in the radio band than the black holes
\cite{2006MNRAS.366...79M}, it is only with the advent of the SKA
precursors that we finally have the potential to map out the radio-X-ray
plane for these sources (while the X-rays are just as bright so do not
need new facilities). On any given day there will be 1-10 sources of
interest to monitor, typically at $0.1 - 10$ mJy flux densities, i.e. readily
achievable within an hour (some sources will be much fainter however and
require more time). 

In addition to the slow evolution of sources in the radio-X-ray plane,
X-ray binaries occasionally go into major outbursts, which are associated
with radio flaring \cite{2004MNRAS.355.1105F}. These flaring states are associated with powerful, intermittent jet
production and, when matched with high-sensitivity X-ray observations and
VLBI, are our best hope for understanding the details of the connection
between accretion and jet formation. We note that
because the sources are typically bright but rapidly varying during
these phases, and that it is the comparison with the X-ray data which
provides the physical insights, cadence can trump raw sensitivity. In fact, such observations could sometimes
be undertaken with a minor subarray of MeerKAT, or even with a refurbished KAT-7.


\subsubsection{Tidal disruption events}

Tidal Disruption Events (TDEs) result from the catastrophic
destruction and subsequent accretion of a star by the gravitational
field of a supermassive black hole. They provide our best insight into
the response of a supermassive black hole to a sudden increase in mass
accretion rate. It now seems that this response may be quite different
to the 'steadier' modes that are analogous to X-ray binaries, and yet
this could have been the dominant mode during the phases of strongest
AGN activity at redshifts 2-3. The current rate of discovery of good
TDE candidates is quite low, probably less than 5 per year. The recent
detection of weak radio emission from the 'thermal' TDE ASASSN-14li
\cite{2016Sci...351...62V} suggests that all TDEs produce radio emission,
but that there may be a range of kinetic modes and powers. As with neutron
star (above) and white dwarf (below) accretion, the huge sensitivity
leap with MeerKAT will allow us to probe a significant number of TDE
radio afterglows for the first time (currently five in total).


\subsubsection{Ultraluminous X-ray sources}

ULX are off-centre X-ray sources in galaxies that have luminosities in
excess (in some cases) of the maximum for stellar-mass black holes.
They may be our best route to finding the population of intermediate-mass
black holes (IMBH) which constitute the path from stellar remnants to
supermassive black holes (see also the recent aLIGO results). The best
accreting IMBH candidate to date, HLX X-1, has revealed radio flaring
indicating transient jet production \cite{2012Sci...337..554W}. A second flaring
`HLX' candidate \cite{2015MNRAS.454L..26H} has since been discovered. Flaring
transient ULXs in nearby galaxies also provide a way to probe
Eddington-rate accretion, showing how jets couple to the accretion
flow in a radiation-pressure dominated environment, as was likely
required for the rapid growth of the first quasars.


\subsection{White dwarf accretion}

\subsubsection{Outflows of accretion-powered outbursts of white dwarfs}

Cataclysmic variables (CVs) show violent outbursts similar to those outlined
above for XRBs, but have as an accretor a white dwarf. It is assumed that
outbursts in both classes of objects are triggered by the same disc instability.
The similarity between XRBs and CVs is most prominent for dwarf novae (DNe).
These non-magnetic, disc-accreting white dwarfs show fairly regular outbursts
that are thought to be powered solely by accretion.

Radio emission from CVs had been reported previously but was usually not
reproducible. It had thus been suggested that the radio emission is correlated
with the optical outburst. This has been confirmed by \cite{2008Sci...320.1318K}
with a detailed radio light-curve of SS Cyg using MERLIN and VLA. Following
the repeatable behaviour of SS Cyg, a subsequent survey by South African PhD
student Deanne Coppejans with the JVLA using triggered observations of 5 DNe
in outburst yielded a 100\% detection rate. The triggers have been provided
by amateur astronomers from the AAVSO. Most of these nearby sources (100-260 pc),
show flux densities around 15-60 $\mu$Jy \cite{2016MNRAS.463.2229C}. 
In addition to these transient CVs, persistent CVs of both magnetic
and non-magnetic nature show radio emission \cite{2015MNRAS.451.3801C}.

MeerLICHT will greatly enhance the study of accreting white dwarfs.
Radio observations will trace the outflowing material of CVs while the
optical observations will provide us real time information on the current
state of the accretion disc. MeerLICHT's 1 minute cadence will yield
optical fluxes and colours as well as variability information. Unlike in
XRBs, the accretion disc in CVs is visible at optical and UV wavelengths.
Thus, any night-time radio observation of a CV will allow us to study both
the outflow and the accretion flow for the first time truly simultaneously.
A MeerKAT in-depth study of a large sample of southern CVs, sampling accreting
white dwarfs with a range of different donor stars (from giants to white dwarfs)
will determine the radio properties of the population of accreting white dwarfs
and will allow for a direct comparison of the outflow properties of
non-relativistic accretors with the relativistic black hole and neutron star XRBs. 


\subsubsection{Outflows in thermonuclear eruptions on white dwarfs}

Radio observations can provide the best estimates of the fundamental
nova parameters, e.g. distance (via expansion parallax from high resolution
imaging), ejected mass and kinetic energy (via modelling radio light curves)
and ejecta geometry (via imaging and light curve modelling). The ENOVA
program on the JVLA has demonstrated that low frequency radio observations
like those that will be made with the initial L-band of MeerKAT, will still provide new insights into
the outflow properties of novae as shown by the recent results on T Pyx
\cite{2014ApJ...785...78N} and V1723 Aql \cite{2016MNRAS.457..887W}. It is now well
understood that in most novae the mass is not ejected in spherically
symmetric fashion, but instead shaped in high velocity, strongly
collimated bipolar outflows \cite{2009ApJ...706..738W,2014Natur.514..339C}.

The need for a southern program can be demonstrated by the recent nova
V1369 Cen (2013), which has not been properly covered in the radio,
even though it was detected by Fermi/LAT and even visible to the naked
eye. KAT-7 observations were attempted with an upper limit of 6 mJy at
1.8 GHz (3 sigma). Through a dedicated ToO programme of Galactic novae
with MeerKAT - in combination with simultaneous X-ray and
optical/near-infrared observations - we can make critical advances in
our understanding of nova ejecta (mass, energetics, nature of the outflow).


\subsection{Cosmic explosions}

\begin{figure}
\includegraphics[width=0.4\textwidth]{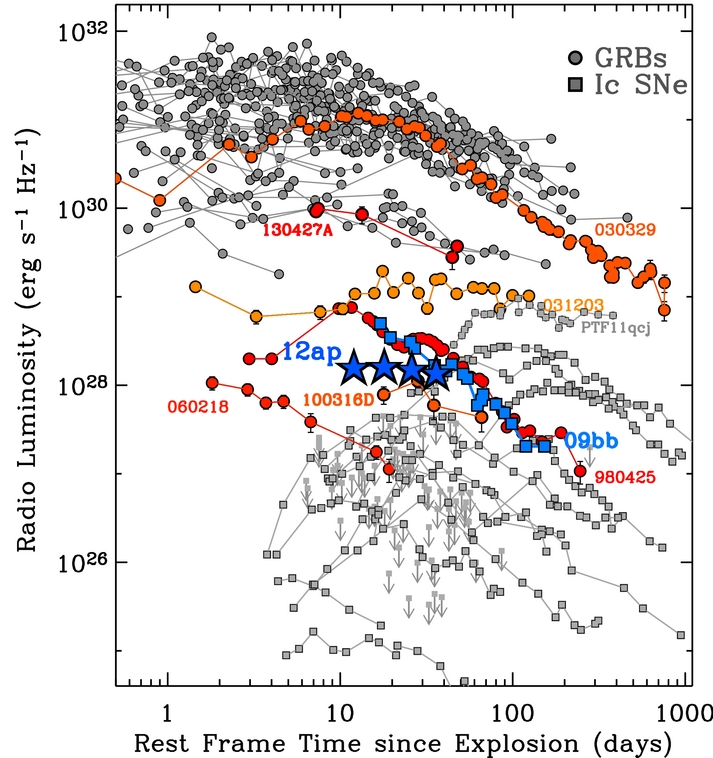}
\includegraphics[width=0.6\textwidth]{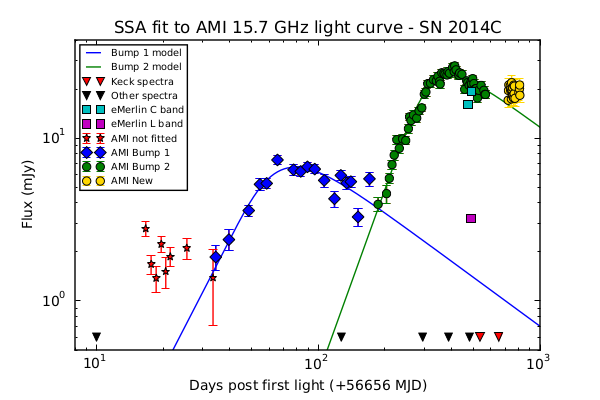}
\caption{Left panel: The zoo of radio light curves of stellar explosions from classical
  supernovae (bottom of panel) to GRB afterglows (top of panel), representing a progression
  from sub- to highly-relativistic outflows. From \cite{2014ApJ...797..107M}.
  Right panel: The extraordinary radio light curve of SN2014 (AMI-LA data). Three,
  possibly four, peaks have been observed in the radio light curve of this source so far,
  indicating interactions of the explosive ejecta with a number of discrete shells,
  providing a unique probe of the last decades (SN ejecta are an order of magnitude
  faster than the winds which created the shells) of pre-supernova evolution and
  mass loss of the most massive stars \cite{2017MNRAS.466.3648A}.}
\label{fig2}
\end{figure}

The endpoints of the lives of the most massive stars in the universe
are the core collapse supernovae (CCSNe) and gamma-ray bursts (GRBs),
the most powerful explosions in the universe. In recent years a picture
is beginning to emerge of a continuum of kinetic powers and Lorentz factors,
with relativistic, engine-powered CCSNe bridging the gap between classical
CCSNe and GRBs (Fig \ref{fig2}, left panel). Our targetted programme is augmented
by commensal studies.

\subsubsection{Gamma-ray bursts}

At all redshifts (currently to $z \sim 8$) half of GRB afterglows are detected
in radio with present techniques. This means we can use them as probes
of the distant universe, and the star formation process through cosmic
history. The special class of short GRBs (duration under 2 s) are thought
to result from compact-object mergers and thus also the prime candidates
for EM counterparts to gravitational wave events (Sect 2.5).

The connection between (long) GRBs and SNe is still uncertain, and needs
to be further explored. According to the best available hypothesis, what
makes a GRB different is that it signals the birth of a black hole from
a fast-rotating star, which unlike a SN causes significant energy to be
emitted as ultra-relativistic jets, giving rise to the spectacularly bright
flashes of gamma rays and afterglows lasting days (X rays) to weeks (optical)
to years (radio). GRBs are exceedingly rare: only about 1/Myr/galaxy, 10,000
times rarer than SNe. This predicts, among other things, that for every GRB
we see, there are 100 whose collimated outflow is not aimed at us. These should
be visible as SNe of type Ib/c, for which at late times, when the GRB afterglow
becomes non-relativistic and unbeamed, we see a GRB-like radio afterglow.
A few recent type Ib/c SNe have shown that a (mildly) relativistic component o
f the SN ejecta can sometimes already be observed in the radio a few days
after the initial stellar explosion, while a gamma-ray signal has not been detected.

Targeted MeerKAT follow-up of GRBs will help us constrain the GRB physics,
determine the true total kinetic energy released, and understand relativistic
shocks and particle acceleration in them. Especially for short GRBs, where less
than 5 radio afterglows have been seen, this will help greatly in improving our
physical understanding of them. Similar follow-up of SNe (Sect. 2.3.2) will
allow us to compare this with the non-relativistic version of the same process,
and possibly also find the GRB-SN connection through detecting GRBs that were
not initially collimated towards us.

Commensal observations in deep fields, such as the LADUMA, MHONGOOSE and Fornax
fields, will allow the detection of a more complete sample of radio events,
possibly including highly obscured ones, and better establish the population
ratios. Tying down the flux and spectrum of GRBs and SNe in the MeerKAT radio
bands will allow us to determine the complete physical parameters of the
explosion, which has only been possible for a few out of many hundreds of
GRBs with afterglows so far. Once we can compare the behaviour in the L band
with possible other bands of MeerKAT, this will become all the more powerful.


\subsubsection{Core collapse supernovae}

Many CCSNe are not observed at radio wavelengths or, if observed, only
followed at radio if there is a detection early after its explosion or
optical detection (a few days to a few weeks). The sensitivity of MeerKAT
will allow us to follow a statistically meaningful number of exploding
CCSNe per year within 40 Mpc with a modest amount of MeerKAT time, permitting
us to also obtain the radio luminosity function of CCSNe. Some individual
CCSNe result in unique views of the final decades of evolution of the most
massive stars (e.g. SN2014C, right panel of Fig 2).

The ideal cadence for the commensal
detections of CCSNe is that the same fields are visited 3-5 times per year
(LADUMA, MHONGOOSE). Assuming a maximum effective distance of 150 Mpc to
discover CCSNe, we estimate that ThunderKAT could yield between 20 and 50
blind CCSNe discoveries, every year.


\subsubsection{Type Ia supernovae with MeerKAT}

Type Ia SNe, the end-products of white dwarfs, have been used to make
the essential discovery of the accelerating expansion of the universe,
yet we do not know what their progenitors are: a C+O white dwarf and a
non-degenerate star (single-degenerate scenario), or two white dwarfs
(double-degenerate scenario). Radio observations can discriminate between
the progenitor models of SNe Ia. For example, in all scenarios with mass
transfer from a companion, a significant amount of circumstellar gas is
expected, and therefore a shock is bound to form when the supernova ejecta
are expelled. The situation would then be very similar to circumstellar
interaction in core-collapse SNe (see Sect. 2.3.1), where the interaction
of the blast wave from the supernova with its circumstellar medium results
in strong radio and X-ray emission.  On the other hand, the double-degenerate
scenario will not give rise to any circumstellar medium close to the
progenitor system, and hence essentially no prompt radio emission is expected.

We aim to perform deep ToO observations of up to two nearby (D $<$ 20 Mpc) Type Ia
SNe per year with ThunderKAT, aimed at making the first ever discovery of
radio emission from a Type Ia SN (the estimated rate for SN Ia within this
volume is 1 $\pm$ 1 per year). We remark that a detection would represent
a breakthrough in our understanding of Type Ia SN progenitors, given the
existing contraints on SN2011fe \cite{2012ApJ...746...21H} and on SN2014J
\cite{2014ApJ...792...38P}, and would validate the single-degenerate scenario.
On the other hand, systematic non-detections will be so constraining that it
will rule out essentially most, or all, single-degenerate scenarios, leaving
the double-degenerate scenario as the only viable one.

We need the observations to be scheduled as soon as the SN candidate is
confirmed as a type Ia SN, and to be done at the highest available frequency
of the array, aiming at a rms of around or better than 1 $\mu$Jy/beam. 


\subsection{Fast and coherent transients}

\subsubsection{Fast radio bursts}

The nature of Fast Radio Bursts (FRBs) remains uncertain, and the largest
fraction of these discovered and studied with MeerKAT are going to be done
so via the TRAPUM/MeerTRAP projects. Nevertheless, for a telescope as sensitive
as MeerKAT, it is entirely possible to discover a very bright FRB in our commensal
short timescale imaging, and we anticipate finding and localising a significant
number of FRBs this way via our commensal observations. Fast imaging of
observations commensal with TRAPUM will be a joint TRAPUM-ThunderKAT collaboration.

However, as well as the short coherent burst there remains the possibility that there is
a synchrotron afterglow of the event \cite{2016Natur.530..453K}, and subsequent literature,
and/or that some events repeat \cite{2016Natur.531..202S} and may be associated with faint
continuous sources. 


\subsection{Gravitational wave sources}

\subsubsection{Gravitational wave events and electro-magnetic counterparts}

The recent direct detections of gravitational waves (GWs) from extragalactic
binary black hole mergers by the advanced LIGO (aLIGO) detector has opened a
new window into the Universe \cite{2016PhRvL.116f1102A}. This discovery, together with the expected
increase in the aLIGO sensitivity, and the addition of VIRGO detectors,
suggests that the coming years will see many compact binary star mergers.
The current localization of these GW sources is several hundred square degrees
(90\% containment probability), but this will likely improve to 300 deg$^2$  and
100 deg$^2$ by 2018 and by 2022, respectively. Finding the electromagnetic (EM)
counterparts of GW sources is crucial for increasing the science yield from
the aLIGO/VIRGO detections, such as 1) lifting degeneracies associated
with the binary parameters inferred from the GW signal, 2) reducing the
SNR for a significant GW detection by aLIGO/VIRGO, and 3) providing arcsec
localization and identifying the precise merger redshift, thus setting/testing
the energy and distance scales.

Compact binary mergers containing neutron stars, i.e. NS-NS/NS-BH systems,
are expected to produce EM counterparts. With MeerKAT, the goal will be to
study the radio afterglows, which probe the post-merger outflows and the
(shocked) circum-merger environments. These afterglows are expected to be weak,
10s to 100s of $\mu$Jy, and expected to peak on weeks -- months timescales after
the merger and detection of the GW signal. Although faint, such afterglows will
probably be the only means of finding the EM counterparts of mergers located in
dusty galaxies, where the optical kilonova could be highly extincted. With
MeerLICHT, we will hunt for the blue (red) kilonova signature that is expected
on day (week) timescales. We envisage the MeerKAT follow up observations to
be carried out in four epochs logarithmically spaced between  1 day and
1 year post-merger. Subsequent monitoring/follow up will be necessary only
for the radio transients identified in these epochs.

\section{Commensal transient searches}

In parallel with the targetted programmes outlined above, we will search all
incoming data streams in the image plane for transients. We will aim to do
this on timescales from $\sim 1$ sec upwards, at approximately logarithmic time
intervals. For example, for a 'typical' 4hr observation we may search for
transients at timescales of 1, 10, 100 and 1000 seconds. These searches
will be carried out in near real time and will probably operate alongside
the beam-formed transient searches associated with the TRAPUM and MeerTRAP
projects. 

By 'piggybacking' on the other approved LSP surveys, from wide to very deep
fields, we will perform the most extensive blind image-plane radio search for
transients ever undertaken. In some of our core areas, e.g. SNe, we can
confidently predict that the commensal results will augment the science, but
the true value of the integrated commensal survey lies in providing a
comprehensive understanding of the radio sky variability at all timescales
and flux levels. We note that for L-band and lower frequency observations of
transients it would be our intention to search for (inter)galactic HI absorption features.

We would like to stress that this aspect of the programme is important in that
it is the major `exploration of the unknown' aspect of the MeerKAT LSP, and a
very important preparatory study for the SKA. However, the clearer knowledge we
have gathered over the past half a decade shows that it is with targetted
programmes such as those we have outlined above, that we will get the
guaranteed high-profile science.

\section{Comments on observing strategy}

Within the ThunderKAT programme, we will need to use a range of different observing modes, including different speeds of ToO (in terms of response time) interleaved with regular observations (some of which may be regular monitoring of a significant number of point sources).

\subsection{Commensality with other Large Survey Projects}

Following the original MeerKAT LSP awards, we agreed with the other LSPs that
we could search their data commensally for transients (with some appropriate
'reward' in terms of authorship). For the revised 2016 LSP plans we note:
\begin{itemize}
\item{MeerTIME / TRAPUM / MeerGAL - These projects will search extensively in
the galactic plane and globular clusters, which is good for us for searches for
galactic transients;}
  \item{Fornax - This project will observe over two 4.5-hr epochs
    per pointing across the Fornax survey area;}
    \item{MHONGOOSE - This LSP will provide $6 - 8$
pointings per deep field, providing an excellent resource to look for faint
(extragalactic) transients over time scales of weeks to months;}
      \item{LADUMA - This
project provides the longest baseline for transient detections on a single
pointing over time scales of weeks to months to years;}
        \item{MIGHTEE - We will provide
          a time-variable sky model to MIGHTEE.}
          \end{itemize}

\subsection{MeerLICHT and the need for some night time observations}

All of the ThunderKAT science benefits from having simultaneous
optical data during the MeerKAT observations. For transients discovered in
commensal searches, the simultaneous optical flux and radio flux density measurements
of an astrophysical transient allow an initial classification (Stewart et al. 2016,
in prep) which will assist in allocating and prioritising multi-wavelength follow-up
resources.  In this regard, ThunderKAT is already part of a consolidated multi-year
South African-led program for rapid follow up of various classes of transients on the
Southern African Large Telescope (SALT).

\section*{Acknowledgements}

This proceeding is an adapted version of the 2016 updated ThunderKAT project plan submitted
to SKA South Africa in June 2016, ahead of the first light images of MeerKAT-16 released on 30 June 2016.
The authorship of this manuscript reflects the full ThunderKAT membership at the time of submission
of these proceedings, and are presented in the following order: the two principal investigators
of ThunderKAT (Fender and Woudt), and alphabetical list of the work group leaders, followed by a second
alphabetical list of ThunderKAT co-investigators. 
RF acknowleges partial support from the European Research Council.
PAW acknowledges the University of Cape Town and the South
African National Research Foundation for financial support. 

\bibliographystyle{JHEP}
\bibliography{ThunderKAT2016_PW17}

\end{document}